\documentclass[twocolumn,trackchanges]{aastex631}

\usepackage{mathptmx}
\usepackage[T1]{fontenc}
\usepackage{ae,aecompl}
\usepackage{float}
\usepackage{graphicx}	
\usepackage{amsmath}	
\usepackage{amssymb}	
\usepackage{svg}

\begin{document}

\title{Charged Clouds of Ionized Gas Emerge from Tribocharging Grains}

\author[0009-0004-2042-780X]{Patrick Hock}
\affiliation{University of Duisburg-Essen, Faculty of Physics,
Lotharstr. 1-21, 47057 Duisburg, Germany}

\author[0000-0003-4468-4937]{Jens Teiser}
\affiliation{University of Duisburg-Essen, Faculty of Physics,
Lotharstr. 1-21, 47057 Duisburg, Germany}

\author[0000-0002-7962-4961]{Gerhard Wurm}
\affiliation{University of Duisburg-Essen, Faculty of Physics,
Lotharstr. 1-21, 47057 Duisburg, Germany}

\begin{abstract}
\noindent If two solid particles collide, charge is exchanged. However, this transfer is not restricted to the surface of the particle. Ions are also  {dispersed} into the environment. They form a charge cloud around the particle. In this way, all particle-laden atmospheres from volcanic plumes on Earth over exoplanet atmospheres to protoplanetary disks might be subject to gas-phase ionization by means of particle collisions.
In laboratory experiments, we quantify the amount of ions produced in a collision of glass beads with 2.8\,mm diameter.
We extract the ions by applying an external electrostatic field and measuring the generated current. The ions are detected at all the pressures studied, i.e. from 0.3\,mbar to 100\,mbar. However, the ionization rate peaks at about 1\,mbar. Scaled to individual bouncing collisions, charge as high as
$\sim$ 1\,pC of each polarity was detected. This implies collisions of grains can be a significant source of ions in various atmospheres.

\end{abstract}


\section{Introduction}

In atmospheric electricity, several processes are discussed as sources of gas-phase ions. 
Ions in the atmosphere are generated by photo-ionization or collisions with energetic particles originating as cosmic rays or from radioactive decay \citep{Viggiano1993In}. Cosmic ray ionization is ubiquitous in Earth's atmosphere \citep{Dartnell2011, Harrison2003Ion‐aerosol‐cloud} and becomes the dominant ionization mechanism at an altitude greater than 1\,km \citep{Hoppel1986}. 
Charges also enter the atmosphere by means of charged solid grains
during the formation of thunderstorms \citep{Mason1988, Saunders2008} or during volcanic eruptions \citep{CIMARELLI2022107449, Mendez2021, Nicoll2019}. Collisional charging is also important for electrostatic activities on other planets \citep{Kok2009, Harper2018}. Tribocharging also influences the formation of planets within protoplanetary "atmospheres" \citep{Muranushi2010, Desch2000, Steinpilz2020, Teiser2025}.

Models of tribocharging regularly consider a direct exchange of charge at the contact between two solids. Therefore, a bouncing grain has a different net charge after a collision. If this process is biased, for example, due to grain size and if different grain sizes are spatially separated, large electric fields can build up and cause lightning \citep{Mason1988, Cimarelli2014, Desch2000}. This produces atmospheric ions.
However, charges can already become entrained in the environment on a much smaller scale.
Indicators of small-scale ionization of the gas phase surrounding tribocharging grains can be found in \citet{Krauss2003}. They observed small-scale discharges in a stirred granular medium.
\citet{Schoenau2021} observed corona-like discharge glow on the centimeter scale generated by a vibrated granular medium at low pressure.

\citet{Jungmann2021} quantitatively determined the net charge on two submillimeter glass particles before and after a collision in microgravity experiments in an atmosphere of 1\,bar. They found that the charge on the particles was not conserved.  Although an indirect measurement, this observation suggests that some charge was transferred to the surrounding gas phase. \citet{Penner2024} directly measured the free ions generated during tribocharging in a vibrating granular medium. The ionized gas was extracted by forced convection. The ions were detected within a cylindrical capacitor. This measurement method is similar to atmospheric ion measurements with a Gerdien tube (e.g., \citet{Aplin2000}).

\subsection{Ions per collision}

The amount of ions generated in collisions will depend on a number of environmental parameters and particle parameters. Among those might be ambient pressure, water content within the atmosphere, water on particle surfaces, particle material, particle size, collision velocities and collision history, e.g. preexisting charges on the surfaces. However, to pin down the importance of these and other, maybe yet unknown parameters, a basic measurement is needed that is well constrained.

One of the most basic quantities is the amount of ions of a given polarity that are ejected in a single collision between two particles. \citet{Penner2024} e.g. deduced a maximum charge of 200\,pC at 1\,mbar ambient pressure on 900\,$\rm \mu m$ basalt beads from their measurements. These were not single collision experiments though, but collisions occurred in a dense particle bed that was vibrated. The ions were transported by gas flow into a capacitor for measurement. They had to assume a number of arbitrary factors that account for the number of grains, the number of collisions per time, the shielding by close proximity of beads within the granular bed, or various sinks of ions on the way toward the measurement cell. 
The initial value reported by Penner et al. (2024) is significant, as it provides the first indication of substantial charge production. However, beyond this, it primarily serves as a proof of concept, with the reported charge representing an initial estimation of the charge generated in a single collision.
In this work we build on these previous experiments.

\section{Experiment}
\begin{figure*}
    \centering
    \includegraphics[width=\textwidth]{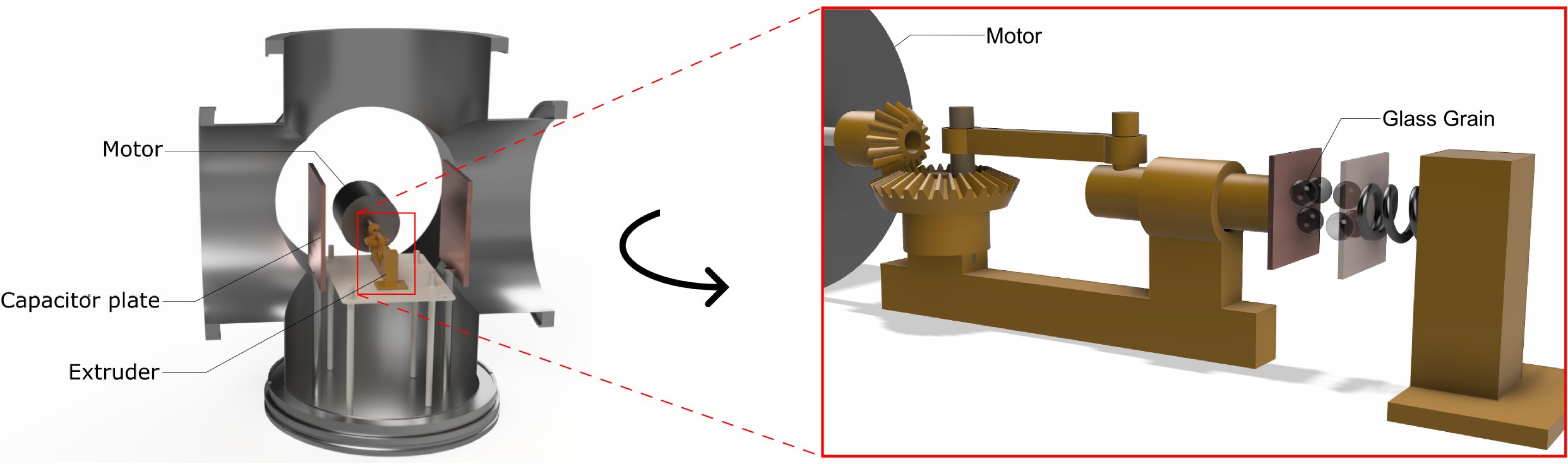}
    \caption{Sketch of the experiment. A mechanism for continuous collision between a small number of glass beads (right) is placed inside a vacuum chamber between two copper electrodes (left).}
    \label{fig:setup}
\end{figure*}

In the experiment by \citet{Penner2024} the collisions and contacts of grains in a dense particle bed depend on the vibration frequency, intensity, free fall, and rebound of ejected particles from the top layer of the granular bed. There are also unseen contact processes in the inner part of the granular bed. In such a setting with thousands of particles in motion, it is essentially unknown how many contacts are made and broken in a given time, and because of different spacings, it is also unknown which contact will provide measurable, free ions at all. To simplify this, we used only six particles where 3 are forced to collide with the others at a fixed frequency. 
Fig. \ref{fig:setup} shows the experimental setup. 

To induce collisions, glass particles are mounted opposite each other on two small copper plates, respectively, as illustrated in fig. \ref{fig:setup}.  These particles are arranged in a triangular configuration. To limit the accelerations during contact, one plate is attached to a spring.  The second particle holder is connected to a motorized linear stage to allow controlled motion. Additionally, one particle setting is rotated by 30 degrees around the barycenter to increase the reliability of collisions.
The square copper bases to which the glass grains are glued with epoxy resin have a side length of 1\,cm and are positioned 1.4\,cm apart at a maximum distance. The rotation of the gear mechanism induces a sinusoidal displacement of one base relative to the other with an amplitude of 1.5\,cm. With the glass particles in between, contact is made about halfway through each oscillation but the spring mounting prevents the system from getting stuck.
Variations in the collision frequency suggest that 3\,Hz to 4\,Hz provide the largest and most stable signals, therefore we use this frequency. The collision velocity is then on the order of 0.1\,m/s.
The dependency on frequency suggests that the strength of collisions is a significant parameter. However, collision velocities and especially accelerations or contact sizes could not be controlled in detail with the given setup.  However, the glass particles were regularly inspected with an optical microscope to exclude any damage on the microscale after collisions. We therefore consider the particles to bounce off each other without generating glass fragments.

As particles for the current study we use soda-lime glass particles with a diameter of $d = 2.8 \pm 0.2$\,mm from \textit{Whitehouse Scientific}. 
The particles were initially cleaned with isopropanol, but were not treated or stored specifically otherwise. Like usual glass, we consider the particles to be hydrophilic, i.e. some physisorption will take place on the surface if they are subject to atmospheric water vapor in the laboratory.

In \citet{Penner2024} the ions within the gas were transported by forced convection into a cylindrical capacitor. This also introduced some unknowns. The gas has to first pass through the particle bed. The ions might collide with other particle surfaces and be lost. In addition, the ions were tied to a gas flow, which was driven by fans. This has limitations at very low pressure, where fans do not operate well. To avoid this transport, the collisions in this work already occur within the measuring capacitor. We consider the trade-off of an external field to be present as negligible for the process as outlined below. As a capacitor, we used two 13 x 13\,cm square copper electrodes at a distance of 13\,cm apart from each other with the colliding particles in the center. The measurement principle is described below, but, in any case, the external field is supposed to draw free ions toward the capacitor electrodes. As seen in fig. \ref{fig:capacitor}, the electric field applied to the electrodes is perpendicular to the connection lines between the particles. We consider this to be the orientation with the least obstruction for ions.

The whole setup is placed inside a vacuum chamber. The vacuum chamber reduces external disturbances, i.e. noise or external ion sources. It also provides the principal means for studying the pressure dependence. Between individual experiments, the gas in the vacuum chamber was replaced with fresh laboratory air and the pressure was adjusted.

\subsection{Charge measurements}

\begin{figure}[H]
    \centering
    \includegraphics[width= 1\columnwidth]{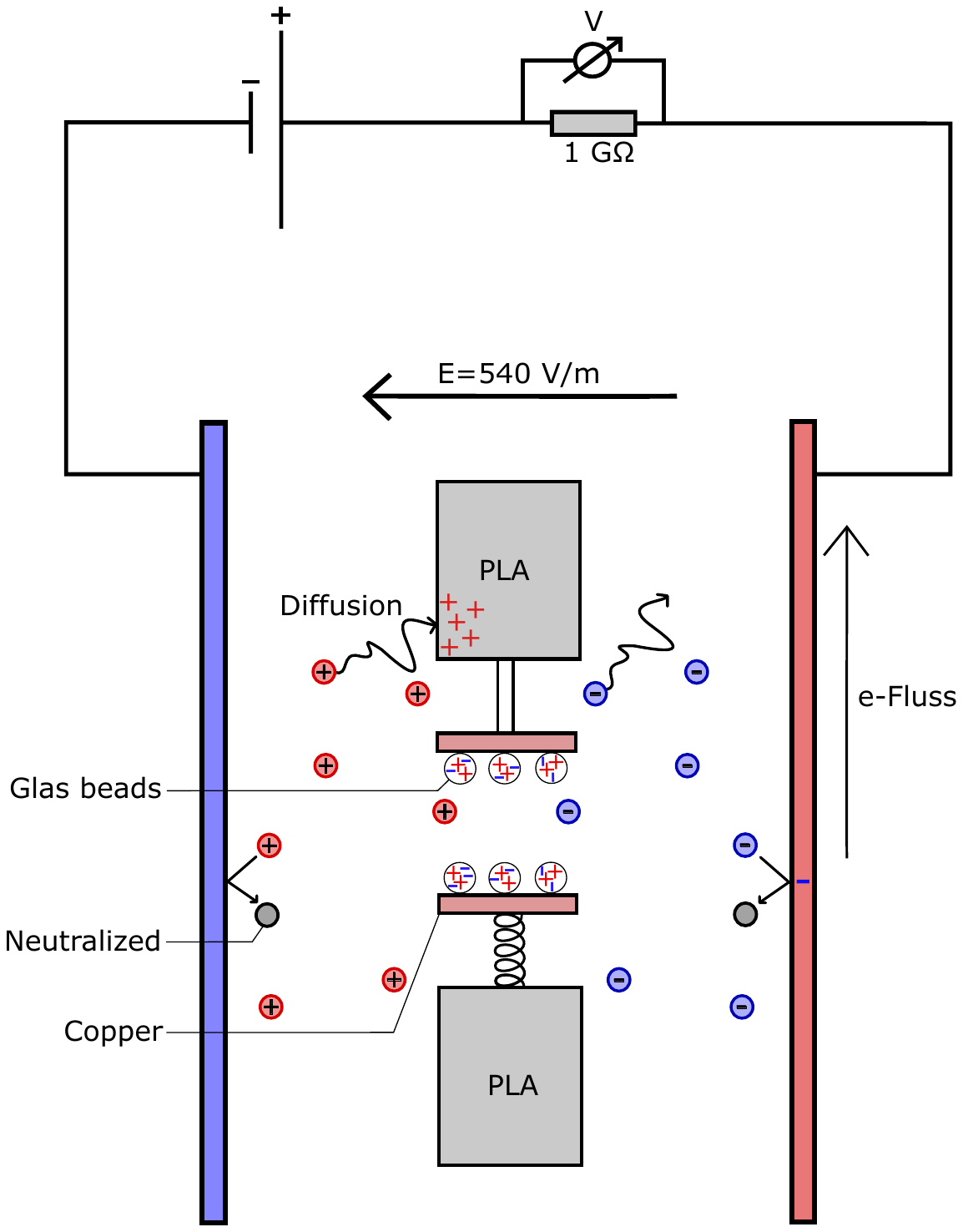}
    \caption{Principle of the charge measurement. The sketch of the capacitor corresponds to the top view and also includes potential ion sinks or recombination sides.}
    \label{fig:capacitor}
\end{figure}

A more detailed sketch of the measurement unit and the processes occurring within is shown in fig. \ref{fig:capacitor}

The copper plates are connected to an external voltage. They provide an electric field acting on the ions between the electrodes. To minimize disturbances associated with a conventional power supply, we utilize a voltage sourced from a series of batteries. The voltage is set regularly to 72\,V. If the collisions are not active, there is no ionization source within the capacitor. Therefore, once the capacitor is charged, no current flows. However, if ions are present, i.e. generated by the collisions, these ions are drawn to the capacitor plates. We assume that each ion hitting an electrode is neutralized. In this way, a current is induced in the circuit.

To measure this current $I$, a resistor of $R =1\, \rm G\Omega$ is placed within the circuit and the voltage drop $U_R = R \cdot I$ along this resistor is measured using a \textit{Keithley 2182} nanovoltmeter. There is a maximum current if the measured voltage is equal to the applied voltage. The closer to this saturation voltage, the smaller the electric field in the capacitor as voltage drop and capacitor voltage have to add up to the applied voltage. However, the measured signal at the resistor is orders of magnitude smaller than the applied voltage. We neglect the back-reaction of the measurements with the electric field of the capacitor.

We do not expect the external field to influence the ionization process. 
As an estimate, one might calculate the voltage of an isolated $r = 1.4 \, \rm mm$ radius particle with a charge of the order of $Q = 1 \, \rm pC$. This is a typical charge set free (see below) and might serve as a quick estimate for the net charge that might reside on a grain. The voltage on the particle would be $U_p = \frac{1}{4\pi\epsilon_0}\frac{Q}{r}= 6.5 \rm \,V$. The electric field on the surface would be $E = U_p / r = 4600 \rm \, V/m$ for homogeneously charged particles. It will be larger at contacts as the surface will be charged highly inhomogeneously \citep{Onyeagusi2022, Steinpilz2020b}. In fact, experiments by \citet{Schoenau2021} or \citet{Wurm2019} show that local fields can be strong enough to cause small-scale atmospheric breakdown. Because of the unknown surface charge pattern, we do not know how many ions would return to the surface and be adsorbed or neutralized. However, this fraction of ions would also not lead to atmospheric charging in the real world. The number of ions might be different in a different geometry with free grains instead of fixed grains. This estimate shows though that the local electric field on the surface is much larger than the external electric field of the capacitor, which is $E = 40 \rm \, V/m$. We therefore do not expect the external field applied for measurements to influence the charge generation process itself. In fact, \citet{Jungmann2022x} applied electrostatic fields of 80\,kV/m while the glass particles collided with the electrodes. They see no influence on charge exchange. However, strong fields could slowly shift charges on a slightly conductive surface \citep{Jungmann2022x}. Therefore, we would also not increase the applied voltage by orders of magnitude as this could eventually influence the charging process.

In essence, we can detect ions that during or shortly after generation become free ions in the sense that they are far enough away that their motion is no longer determined by their host particles. This initial distance, which ions can gain, depends on the ejection process. This is currently unknown, but we discuss two possible ejection mechanisms below.

Independent of the ejection mechanism, free ions are subject to interaction with the external electric field and the ambient gas. In general, interaction with the gas results in Brownian motion, which is a random walk resulting from collisions with gas molecules. Without an electric field, this diffusion can move the ions to arbitrary places. That means they can stick to all surfaces. If they collide with the insulating surfaces of the collider structure, they might charge these surfaces, which, in turn, might add additional electric fields. These effects are currently unknown. If the ions collide with the conductive and grounded experiment chamber, they could just be neutralized.

\begin{figure}[H]
    \centering
    \includegraphics[width= \columnwidth]{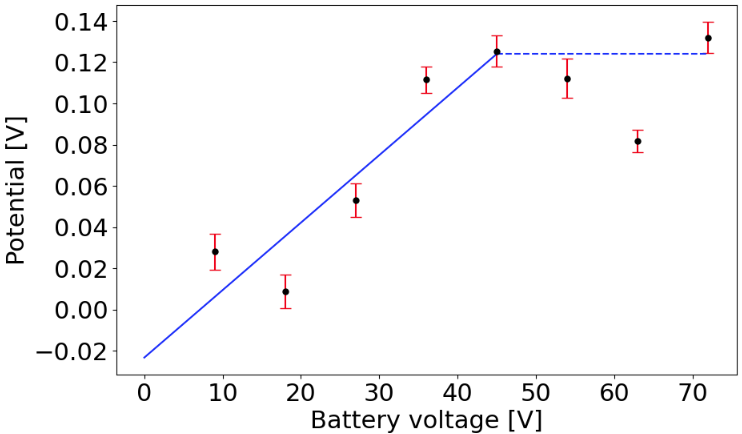}
    \caption{Increase of the measured voltage drop $U_R$ depending on the battery voltage applied to the copper plates for 1 mbar. The blue line is intended to guide the eye, i.e., the rate saturates at about 40\,V.}
    \label{fig:battery}
\end{figure}

In any case, loss due to diffusion can be reduced if the external electric field dominates the motion. This can be estimated as follows. During collisions of the ions with gas molecules, the electric field accelerates the ions and imparts kinetic energy to them. The respective motion is a directed motion toward the capacitor plates. However, collisions of ions with gas molecules randomize the velocity and thermalize the ions. We therefore have two energies, which can be compared. There is the thermal energy $E_t = 1/2 \cdot k\cdot T = 0.025 \rm \, eV$ at room temperature. The directed kinetic energy provided by the electric field $E_k$ can be estimated as follows.  We assume an ambient pressure of 1\,mbar. The mean free path length of gas molecules is about $\Delta x = 70 \,\rm \mu m$. If we assume that the ions are gas ions with one elementary charge, then they gain an energy of $E_k = e \cdot E \cdot \Delta x= 0.04 \rm \, eV $. Therefore, at 1\,mbar, most ions should be collected by the capacitor if 70\,V are applied. The energy of the electric field and the thermal energy balance if 40\,V is applied. We have not discussed the measured signals in detail yet. But if we measure the strength of the signal on the applied voltage we find data shown in fig. 
\ref{fig:battery}.

The voltage increases roughly linearly until saturation is reached at about 40\,V, in agreement with the previous considerations. However, this also implies a situation where, at higher pressures, the maximum voltage currently available is not sufficient to compensate for diffusion, and the efficiency of detecting ions will decrease linearly.

\section{Measured signals}
 
\begin{figure}[H]
    \centering
    \includegraphics[width= \columnwidth]{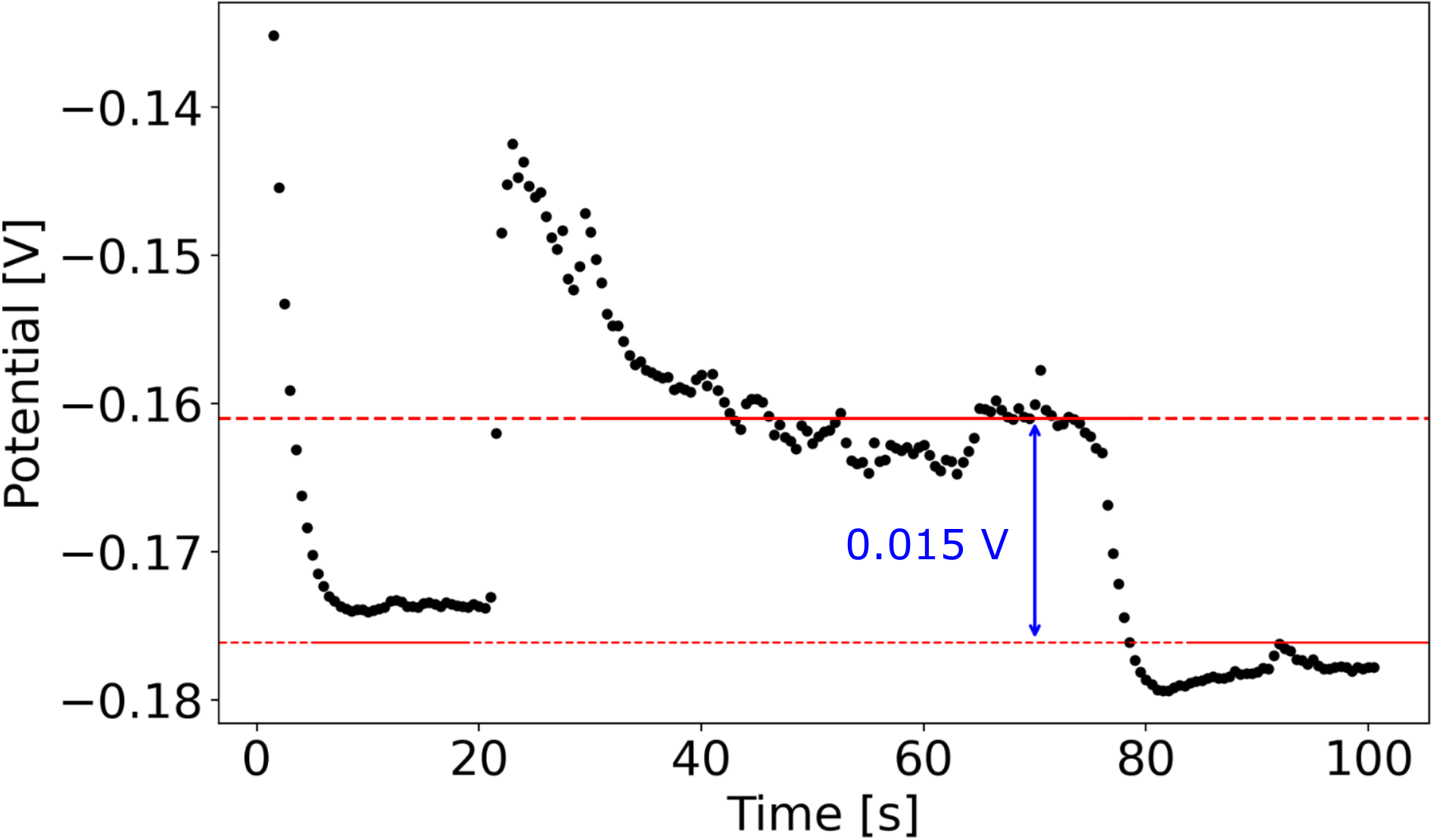}
    \caption{Example of the typical measurement (1\,mbar); At the onset of collisions at about 20\,s, the voltage increases. After some initial variation (20-40\,s), this equilibrates to a constant value (40-75\,s). After collisions stop (75\,s), the voltage drops again to the background value. The continuous red line segments are positioned at the intervals where the mean values are calculated (two segments for background). The lines are extended by dashed lines for visualization.}
    \label{fig:Plateau}
\end{figure}

A typical measurement of the voltage drop $U_R$ on the resistor can be seen in fig. \ref{fig:Plateau}. Polarity can be ignored. Without collisions, there is a voltage bias as the high-ohmic resistor picks up external signals. This varies over time, but the variations are small compared to the ion signals. The background bias serves as the baseline. 

The signal increases as soon as the glass grains collide and drops again when the collisions stop.
No signal is measured if the mechanical system operates but the grains do not make contact.  
The initial signal varies and might be somewhat higher than the equilibrium value, which is reached some time later. We take the latter equilibrium related to the baseline as the core measurement.

The measurements depend on the sample history. 
If measurements are conducted at low pressure over an extended period (several hours), the voltage exhibits a decreasing trend.
We have not yet quantified this process. If the chamber is refilled with air in between measurements, this history is erased and the time dependence is reset in large parts. We suggest that this is indicative of the adsorption of adsorbates on the surface as one mechanism for charging and ion production. The adsorbates would be depleted during repeated collisions by removing molecules from the surfaces. The pressures used in these experiments are relatively high by the standards of surface physics. The replenishment of adsorbates within the vacuum chamber might still not occur effectively. This might also be a question of layer thickness.
Further investigation of the microphysics of this process is beyond the scope of this paper, i.e. it cannot be quantified or interpreted with the current technique applied. 

For the present work, we filled the vacuum chamber with  {laboratory} air after each measurement. This is relevant for situations where particles have time between collisions to equilibrate and where the contact points between particles are not always the same.

\begin{figure*}
    \centering
    \includegraphics[width= \linewidth]{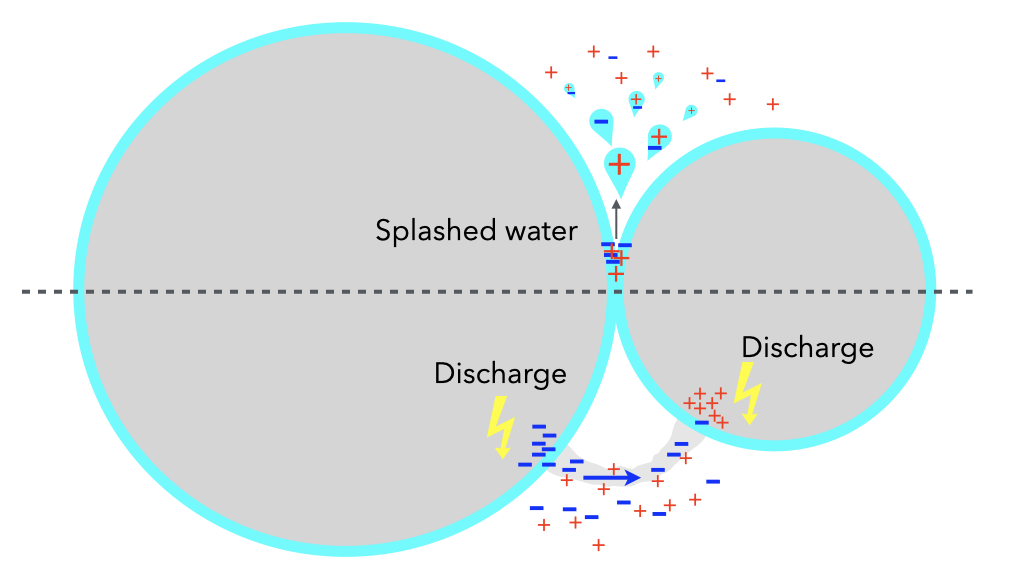}
    \caption{Charge cloud models; top side, water as charge carrier; bottom, atmospheric discharge as ion source}
    \label{fig:chargecloud}
\end{figure*}
\subsection{Pressure dependence}

\begin{figure}[H]
    \centering
   \includegraphics[width= \columnwidth]{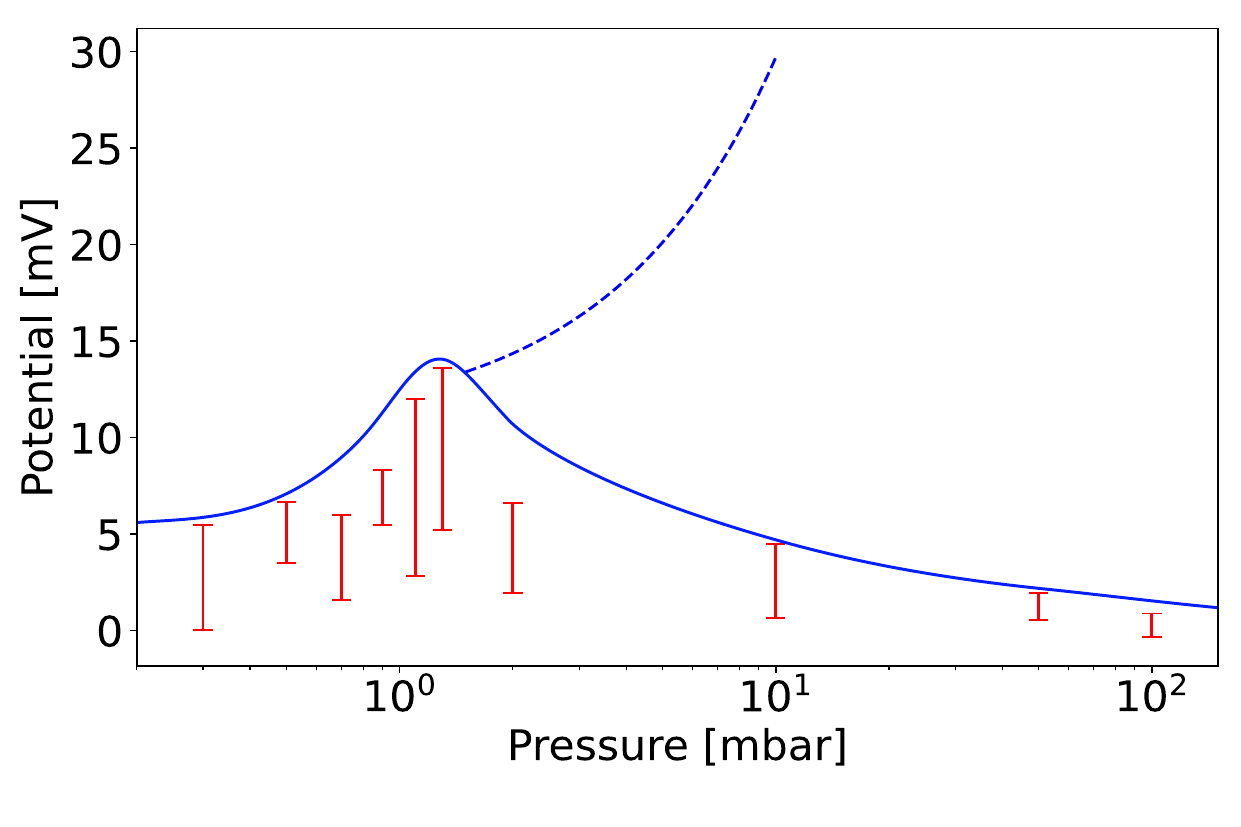}
    \caption{Measured voltage dependence on pressure; error bars in red mark the standard deviation over  {3-5} measurements for each value of pressure. The blue curve is an envelope to visualize the maximum voltages observed. Also added as dashed line is an extrapolation assuming that the sensitivity decreases linearly with pressure.}
    \label{fig:V(p)}
\end{figure}

The works of \citet{Jungmann2021} on the one hand and the works of \citet{Schoenau2021} and \citet{Penner2024} on the other show ionization in different pressure ranges. While the former work sees charges entrained into the ambient gas at atmospheric pressure, the latter two see a maximum at around 1\,mbar pressure, probably related to small-scale discharges \citep{Wurm2019, Becker2022, Cruise2023}.
We also see a peak in our pressure-dependent measurements as seen in fig. \ref{fig:V(p)}.
However, up to the tested pressure of 100\,mbar, there is still a signal related to particle collisions. 
So we expect charge clouds to be generated at all pressures. In view of the discussion on diffusion above, we would also expect that more ions go astray at higher pressure. We therefore add an extrapolation to the data by  assuming a linear decrease in sensitivity with pressure.

\subsection{Charge per collision}
{In their work using vibrated granular beds, \citet{Penner2024} estimated the number of charges generated in a collision, not without some uncertainties. We strive to minimize those uncertainties in our work.}
Many of these uncertainties are not present in our new set-up. 
Most importantly, we only use a very small number of colliding grains at a well-defined collision frequency, which allows us to precisely constrain the rate of charges, $Q$, produced.

The current or charge per time measured is given by the amount of grains colliding $N$, each producing a charge $Q$ with a frequency of the collisions $\nu$ or
\begin{align}
    I=N \cdot  \nu  \cdot Q.
\end{align}
As we measure the current $I$ by the voltage drop $U_R$ at the resistor $R$, we get as individual charge released per contact
\begin{align}
    Q=\dfrac{U}{N\cdot  \nu \cdot R}.
\end{align}
Taking the highest plateau voltage observed of $U_R=0.015$, V at an ambient pressure of 1.3\,mbar, a frequency $\nu = 3$\,Hz, $R = 1\,G\Omega$, and $N=3$ we get $Q=1.68$\,pC per collision.

\section{Discussion}

So far, this is the third experiment in which the ionization of the ambient environment was quantified following work by \citet{Jungmann2021} and \citet{Penner2024}.
\citet{Jungmann2021} estimated large charge losses of about 20\,\% but from the balance of the net charge on two grains before and after a collision in a $CO_{2}$ atmosphere. 
The submillimeter grains in \citet{Jungmann2021} or \citet{Steinpilz2020} charged to the order of a pC. 
The charge varies with size, i.e. increases with size \citep{Wurm2019, Becker2022}. Therefore, somewhat larger charges are expected for the particles used in our study. However, in general, if 20\% is a constant charge loss fraction, ionization of 1\,pC in a collision seems reasonable. 
However, it should be noted that our findings do not relate to a particular net charge. Furthermore, the pressure range and atmosphere between \citet{Jungmann2021} and our work differ. 

It might also be that different ionization mechanisms exist at different pressures. We speculate that the simple ejection of water ions could be one source \citep{Lee2018, Kudin2008}. The discharge ion generation would be  {another} source  (\citet{McCarty2007}). The latter could be related to a peak in pressure dependence.
Both ionization rates might be different, and we could not have sampled the high-pressure (water) rate with our setup correctly, although our time dependence indicates some influence of water.
These two ideas are visualized in fig. \ref{fig:chargecloud}.
Depending on the mechanism, different ions can be charge carriers from positive $O_2^+$ or $H_3O^+$ to negative $O^-$, $OH^-$, or $CO_3^-$
\citep{Skalny2006, Jiang2018}.

Compared to \citet{Penner2024}, our measurement is one to two orders of magnitude smaller than their estimate. Their estimate was based on quite a number of assumptions. It should be pointed out that they used basalt grains, which we (unpublished) regularly found to retain a larger charge by an order of magnitude; so the quantitative values might not be so different after all, but we did not test a material dependence yet.
Since the ion signal is measurable from the moment the particles collide, it might not be necessary to have many collisions first to generate charged particles. Therefore, charge clouds should be easily generated in all colliding systems.

\section{Conclusion}

We improved an earlier experiment by \citet{Penner2024} to measure the charge that is released into the environment upon contact with two mm-sized glass spheres. We found that from about 1\,mm glass beads regularly more than about 1\,pC emerge in a collision, which would form an ion cloud surrounding the grains, if not extracted. This cloud includes both polarities and is not bound to the grains, i.e., it is free to move along with any gas flow.

This ionization process might be dominant in regions with many particle collisions, namely during volcano eruptions, in regions with active dunes or intense aeolian processing, and during thunderstorms. We provide a kind of benchmark value here.
The resulting value should be applied with caution, as numerous additional parameters, which have not been investigated, may influence the outcome.
In any case, whenever tribocharging occurs, it is likely that the surrounding atmosphere (gas) is also efficiently ionized along. 

\section{Acknowledgments}
The project is funded by the Deutsche Forschungsgemeinschaft (DFG, German Research Foundation) under grant 521602700. We appreciate the thorough review by two anonymous referees, which strongly improved the manuscript.

\bibliography{bib}
\bibliographystyle{aasjournal}

\end{document}